\documentclass[a4paper,11pt]{article}
\usepackage{pos}
\usepackage{siunitx}
\usepackage{subfig}
\usepackage{lineno}
\usepackage{enumitem}

\title{Status of the SST Camera for the Cherenkov Telescope Array}
\ShortTitle{Status of the SST Camera for CTA}

\author*[a]{Davide Depaoli}
\affiliation[a]{Max Planck Institute for Nuclear Physics, Saupfercheckweg 1, 69117 Heidelberg, Germany}

\onbehalf{for the CTA SST Project} 

\emailAdd{davide.depaoli@mpi-hd.mpg.de}

\abstract{
  The Cherenkov Telescope Array will be the next generation ground-based gamma ray observatory in the energy range from a few tens of GeV to hundreds of TeV. It will be built on two sites, one for each hemisphere, to cover the entire sky. The observatory will consist of telescopes of three different sizes: large, medium and small, with primary reflectors of 23, 11.5 and 4.3 m in diameter, respectively. The Small-Sized Telescopes (SSTs) will focus on the highest energies; at least 37 (and up to 70) will be deployed at the southern site in Paranal, Chile, covering several square kilometers. They will have a Schwarzschild-Couder dual-mirror design, with a primary reflector of about 4 meters in diameter. This configuration leads to a compact camera, with a diameter of about \qty[]{50}{\centi\metre} and a weight of less than \qty{100}{\kilo\gram}. Its focal plane consists of 2048 Silicon Photomultiplier pixels, each one read independently by a state-of-the-art full waveform readout.
  The camera design is now in the final stage and the first components are being tested. In this contribution we discuss the design choices, and present test results from latest developments.
}
\ConferenceLogo{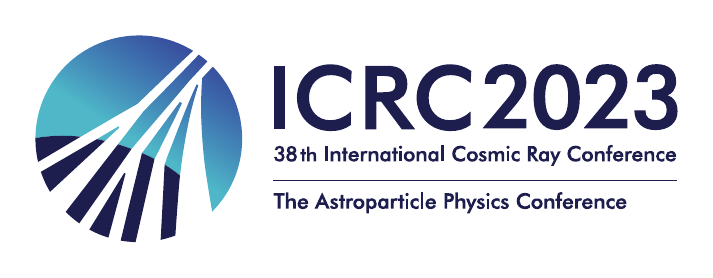}
\FullConference{%
38th International Cosmic Ray Conference (ICRC2023)\\
  26 July - 3 August, 2023\\
  Nagoya, Japan}

\begin{document}
\maketitle
%
%
\section{Introduction}
The Small-Sized Telescopes (SSTs) of the Cherenkov Telescope Array (CTA) are a crucial component of the CTA-South site located in Paranal, Chile, where they will extend the sensitivity of the observatory to the highest energies, from about \qty{1}{\tera\electronvolt} to at least \qty[]{300}{\tera\electronvolt}.
At least 37 (and up to 70) will be built there; because of their large number, their camera should be inexpensive, reliable, and easy to maintain.
During the CTA development phase, different solutions were proposed for the SST camera
\cite{Heller_2017, Pareschi_2019, Zorn_2019, Dumas_2014}.
\par
After an initial development and then a harmonization process, CTA selected the SST technology:
the telescopes will have a Schwarzschild-Couder dual-mirror design, based on the ASTRI structure and on the Compact High Energy Camera in its Silicon Photomultiplier (SiPM) variant (CHEC-S) \cite{White_2022}.
The SST camera is therefore the evolution of the CHEC-S prototype, and its design is being finalized by a team of 11 institutes from different countries, located in Germany, the United Kingdom, the Netherlands, Japan and Australia.
The baseline plan is to deploy at least 37 SSTs at the CTA South site (Paranal, Chile).
%
\section{Design Choices}
The SSTs have a dual-mirror design based on the Schwarzschild-Couder principle, which allows for a small, light and compact camera with a diameter of about \qty[]{50}{\centi\metre} and a weight of less than \qty{100}{\kilo\gram}. The camera will cover a field of view of about \qty[]{9}{\degree} and will use SiPMs as the photosensors, which are robust to high illumination  levels, provide high photon detection efficiency, and are extremely affordable for the size required for the SST.
\begin{figure}[!hbtp]
  \centering
  \includegraphics[width=\linewidth]{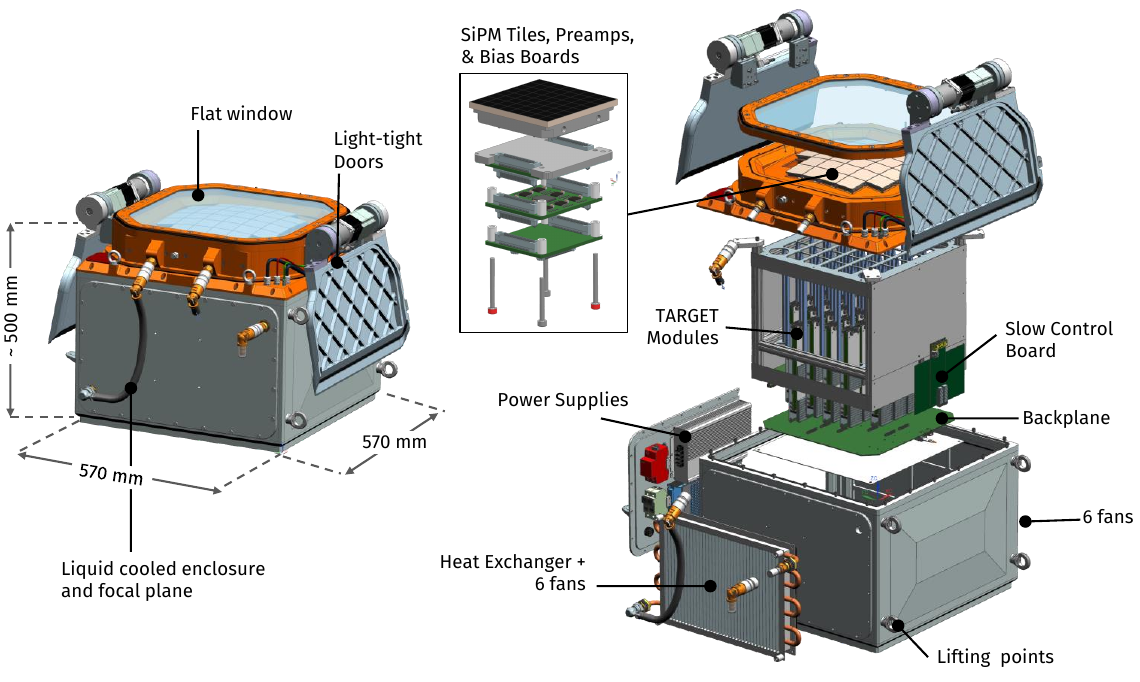}
  \caption[]{Annotated CAD model of the SST camera}
  \label{fig:SST_CameraOverview}
\end{figure}
\par
Figure \ref{fig:SST_CameraOverview} shows the overview design of the SST camera.
The camera is housed in an aluminum enclosure that measures \qtyproduct{570 x 570 x 500}{\milli\meter}; the entire camera weighs approximately \qty[]{90}{\kilo\gram}.
The sensors in the focal plane are protected by a flat window, which has a special layered coating that restricts light outside the \qtyrange[]{290}{550}{\nano\metre} range, to minimize the interference from the night sky background (NSB) photons and maximize the detection of Cherenkov light.
Two light-tight motorized aluminum doors protect the camera during the day.
\par
The camera design is modular and divided into 32 identical modules.
Each module has a tile of 64 SiPMs each, for a camera total of 2048 pixels.
The tiles in the focal plane are arranged to follow the radius of curvature that arises from the telescope optics.
The selected SiPMs are manufactured by Hamamatsu Photonics using the
\qty[]{50}{\micro\metre} LVR3 technology, have an area of \qtyproduct[]{6 x 6}{\milli\metre} and do not have any protective coating (single pixel: S14520-2246, tile part number: S14521-1720).
Tests performed in the collaboration showed a photon detection efficiency of about \qty[]{50}{\percent} at the operating voltage, along with a crosstalk of about \qty[]{5}{\percent}.
\par
Each SiPM tile is thermally connected to an in-house machined liquid-cooled heat sink designed to remove the power dissipated by the sensors and electronics. The temperature is monitored by 4 temperature sensors per tile.
The SiPMs are connected to the Focal Plane Electronics, which is a set of stacked printed circuit boards (PCBs) that provide SiPM bias and signal preamplification.
Each SiPM is individually biased to compensate for different gains in the same tile; the voltage can be set by a \qty[]{16}{\bit} Digital to Analogue Converter (DAC).
The SiPM signals are preamplified by transimpedance amplifiers; a pole-zero cancellation circuit shapes the signals and removes the long signal tail to achieve a Full Width at Half Maximum (FWHM) of approximately \qty[]{10}{\nano\second}.
%
%
The signals are then sent to the the readout electronics, which consists of three interconnected boards per module. 
Based on the TeV Array Readout Electronics with GSa/s sampling and Event Trigger (TARGET) Application Specific Integrated Circuits (ASICs), it performs the first level trigger and the digitization of the signals \cite{Funk_2017}.
All the modules are connected to a Backplane PCB.
\par
The preamplified signals are sent to the TARGET module where they are reamplified and then split into two lines: a fast-signal, and a slow-signal line.
\par
\begin{figure}[!hbtp]
  \centering
  \includegraphics[width=\linewidth]{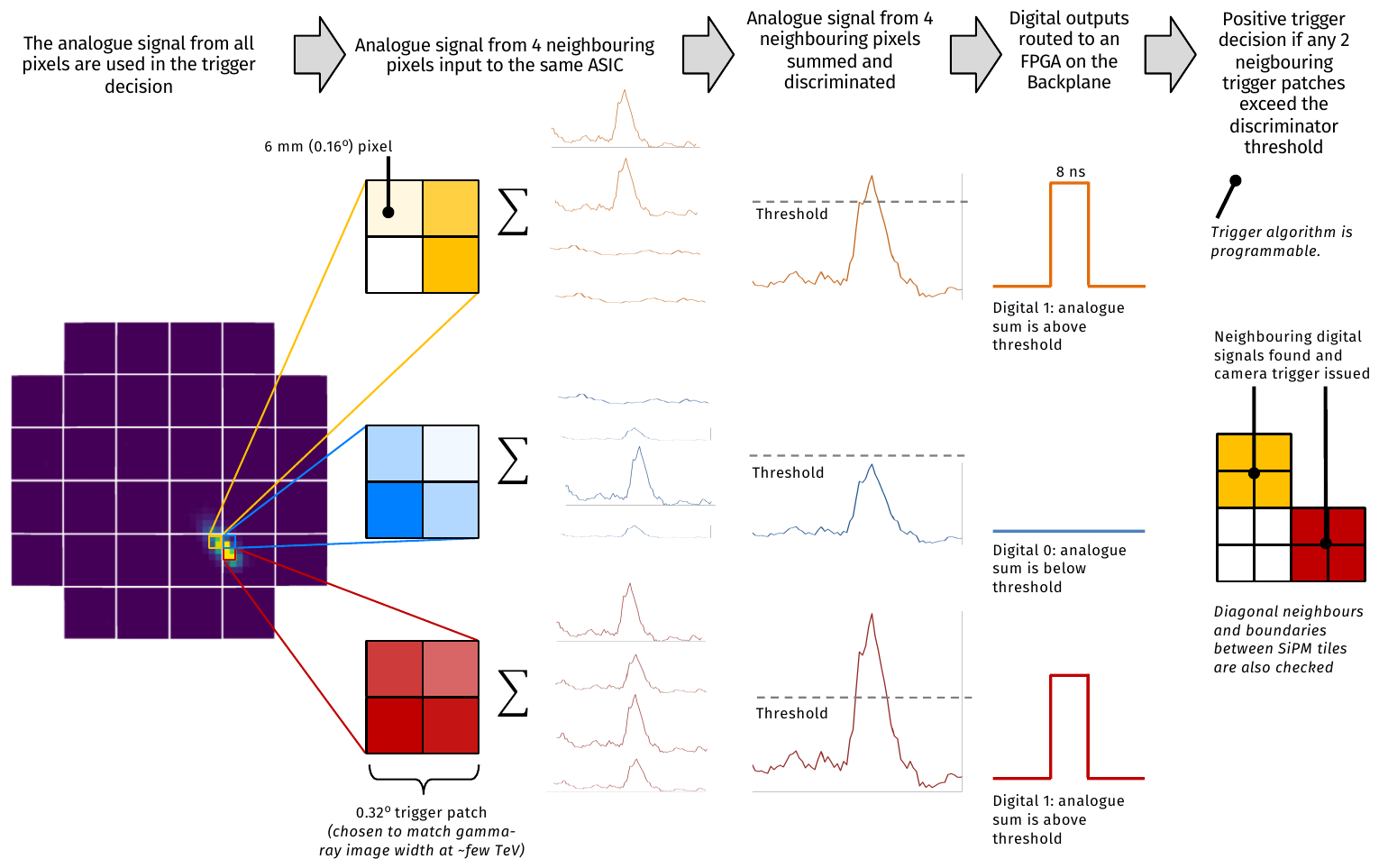}
  \caption[]{Overview of the SST Camera trigger}
  \label{fig:SST_Trigger}
\end{figure}

The fast signal is shaped again and then AC-coupled to the TARGET sampling ASICs (CTC) and the TARGET trigger ASICs (CT5TEA).
Each of these ASICs receives 16 input channels, so each module requires four of them to handle the 64-pixel SiPM tile.
The TARGET-CTC ASICs in the SST camera are configured to sample and digitize the fast signal at \qty[]{1}{GSa/s} in a \qty[]{128}{\nano\second} window (settable in \qty[]{32}{\nano\second} blocks up to \qty[]{512}{\nano\second}). The samples are stored in a capacitor array and then digitized by {12}{bit} Wilkinson ADCs.
The first level trigger is performed by the TARGET-CT5TEA ASICs. A trigger signal is sent when the analogue sum of the fast signals from four adjacent SiPMs (a ``superpixel'') exceeds a configurable threshold.
Each trigger signal from 512 superpixels is sent to the Backplane board, for the determination of the camera-level trigger, configured to require a coincidence between two neighboring superpixels. An overview of the camera trigger is shown in Figure \ref{fig:SST_Trigger}.
Then the signal is sent to the FPGAs hosted in the TARGET modules, which in turn instruct the TARGET-CTCs to digitize the signals (64 per module). These waveforms are then sent to the Backplane, combined into a \qty[]{10}{Gbps} link and then transmitted off-camera via a fiber-optic cable.
\par
In the slow signal line, the DC-coupled signal is integrated by a \qty[]{1}{\micro\second} integrator and then digitized by a \qty[]{16}{bit} ADC; the data is then sent to the FPGAs. The resulting values are used to determine the brightness level for each pixel, which is ideal for measuring NSB levels and for tracking the telescope's pointing using known star locations.
\par
An array-wide White Rabbit system connected to a dedicated Timing Board provides absolute timestamps to each event. 
\par
The camera has a fast and variable intensity illumination system based on LED flashers to allow on-site calibration. The main flasher is located in the center of the secondary mirror and flashes the camera directly. 
Additional flashers are located in the corner of the camera (for indirect illumination by reflection from the secondary mirror) and connected to a scintillating fiber that runs around the inside of the focal plane, below the window (to provide low-light pulses when the door is closed, useful for calibrating the gain of the SiPMs).
\par
The Slow Control Board (based on a general purpose \qty[]{32}{\bit} microcontroller) controls the doors, manages and controls the power distribution, communicates with the LED flashers, communicates with the Backplane, and monitors the camera's internal environment.
\par
An external chiller cools the camera by circulating a cooled liquid through the focal plane plate. Fans inside the camera move the cold air around. The camera is hermetically sealed and humidity is maintained at an acceptable level by a breather desiccator.
%
\section{Development Status and First Measurements}
%
\begin{figure}[!hbtp]
  \centering
  \includegraphics[width=0.9\linewidth]{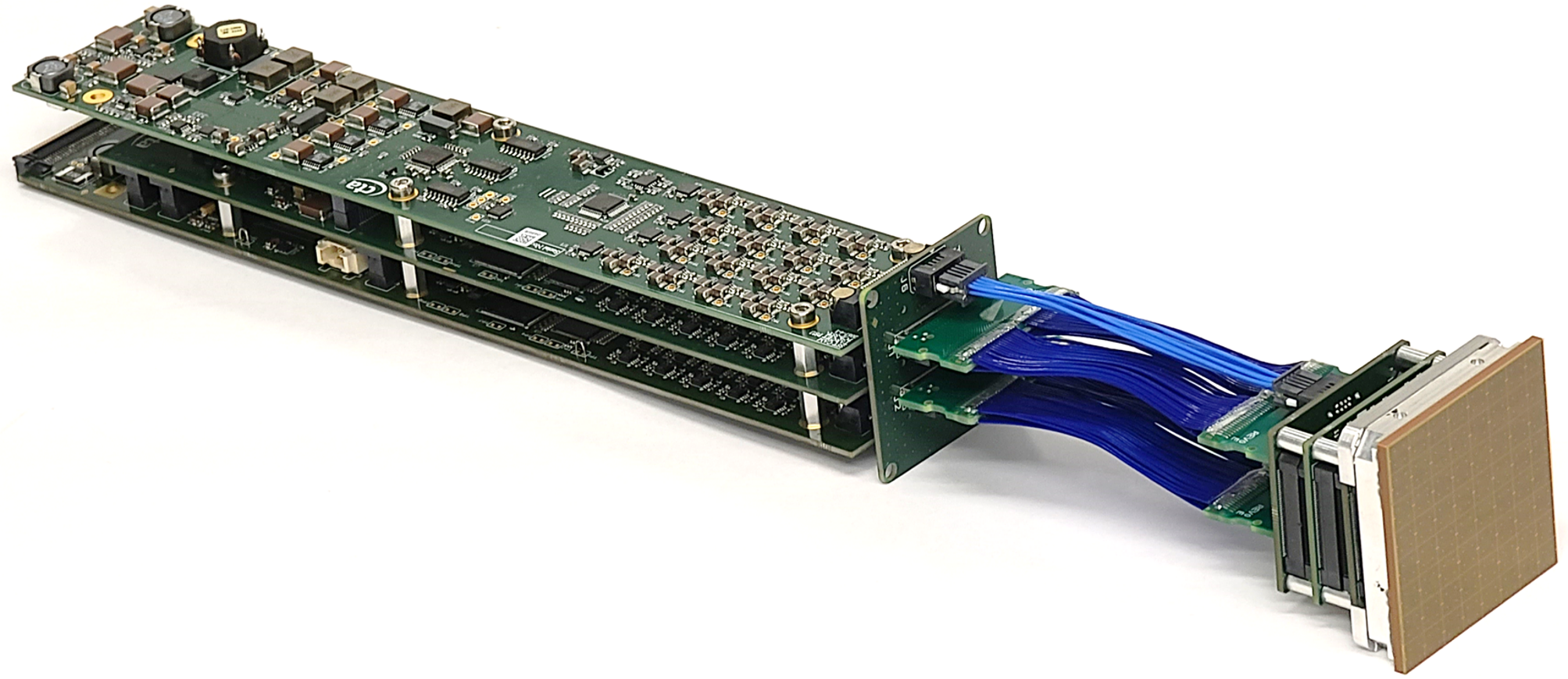}
  \caption[]{Picture of the first SST Camera module }
  \label{fig:SST_Module}
\end{figure}
The design of the SST camera is nearly complete, with only minor adjustments and optimizations remaining. The first complete modules, consisting of TARGET modules, Focal Plane Electronics and SiPMs, have been assembled and are undergoing testing.
The modules are kept in a light-tight box, and the SiPM tile is enclosed in a custom aluminum frame with a UV transparent window connected to a liquid chiller. A picture of the module under test in the University of Leicester laboratories is shown in Figure \ref{fig:SST_Module}.
A focused laser beam mounted on an x-y stage is used to fire one pixel at a time.
During each run, data is stored for every pixel, both fired and unfired.
\par
\begin{figure}
  \centering
  \subfloat[\label{fig:FWHM}]{\includegraphics[width=0.5\columnwidth]{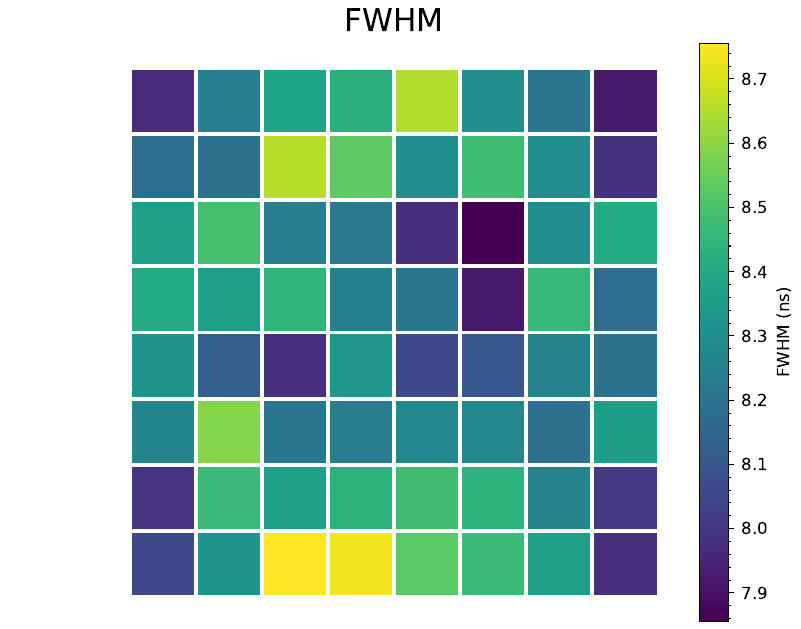}}
  \hfill
  \subfloat[\label{fig:RiseTime}]{\includegraphics[width=0.5\columnwidth]{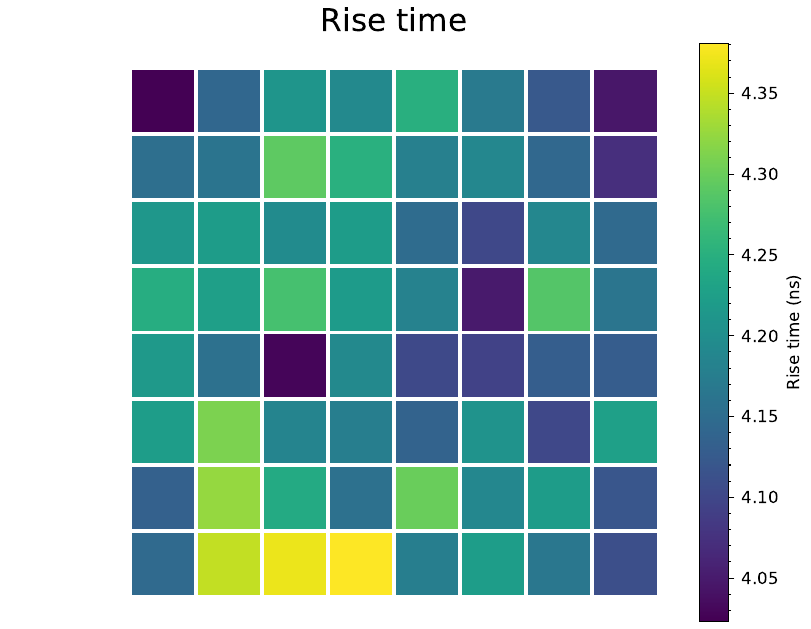}}
  \caption[]{Full width at half maximum (a) and rise time (b) measurement results for the first module produced. Each square is a physical SiPM in the tile.}
  \label{fig:PulseAndTime}
\end{figure}

Figures \ref{fig:FWHM} and \ref{fig:RiseTime} show the results of full width at half maximum and rise time measurements of the signals for each illuminated pixel of a single module.
The response is uniform, with the average FWHM among the tiles being \qty{8.3}{\nano\second} with a standard deviation of \qty{0.2}{\nano\second} (\qty[]{2.5}{\percent}), and the average rise time being \qty{4.2}{\nano\second} with a standard deviation of \qty{0.1}{\nano\second} (\qty[]{1.9}{\percent}).
These preliminary results are in accordance with the design specifications.
\par
\begin{figure}
  \centering
  \subfloat[\label{fig:ct_p2p}]{\includegraphics[width=0.5\columnwidth]{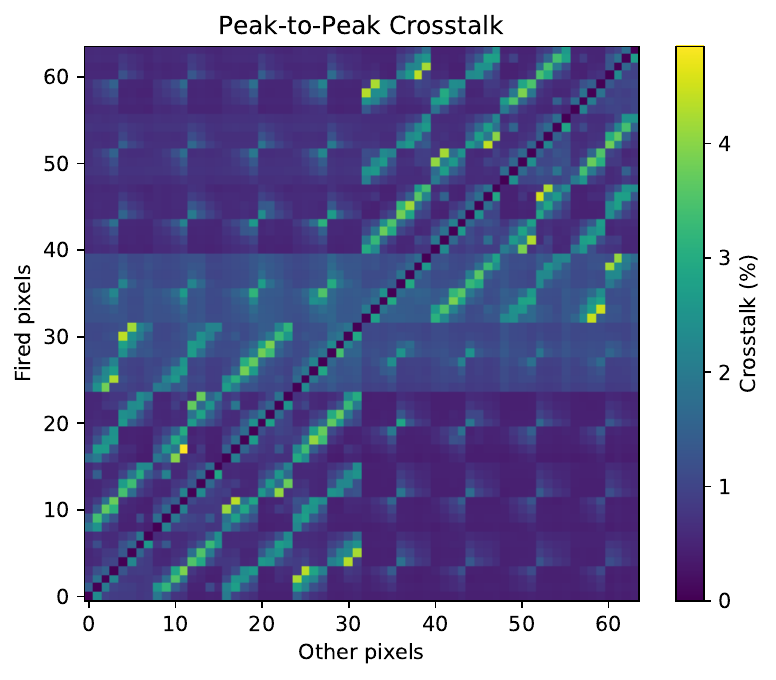}}
  \hfill
  \subfloat[\label{fig:ct_charge}]{\includegraphics[width=0.5\columnwidth]{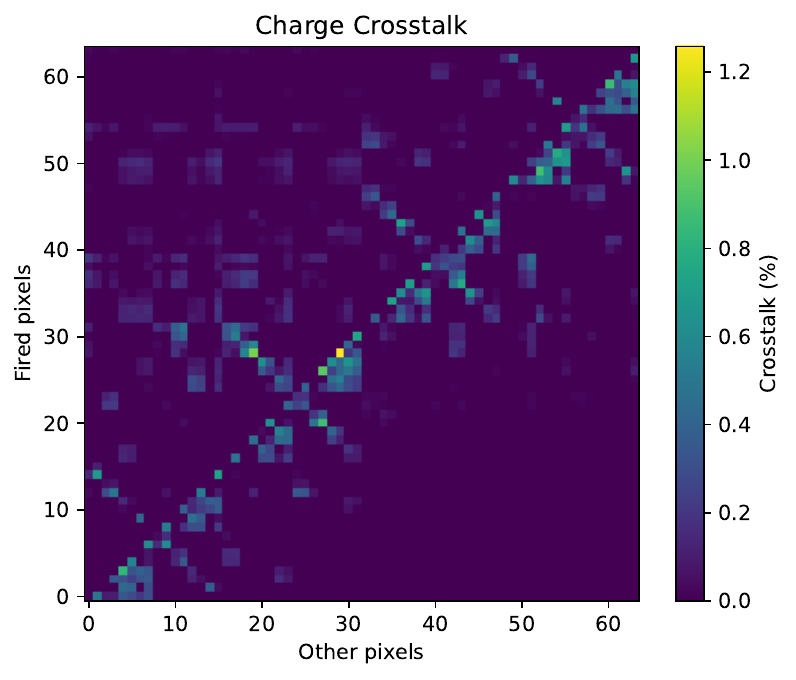}}
  \caption[]{Peak-to-peak (a) and charge (b) crosstalk measurement results for the first module produced}
  \label{fig:ct}
\end{figure}

With this setup, by firing only one pixel with a focused light, it is possible to evaluate the crosstalk between pixels.
The 2D map of the peak-to-peak and charge crosstalk measurements for the module under test is shown respectively in Figure \ref{fig:ct_p2p} and \ref{fig:ct_charge}.
The y-axis shows the fired pixels and the x-axis shows all pixels of the module. For each pixel, the crosstalk is measured as the ratio of the peak-to-peak (or charge) value in the pixel to that in the fired pixel; by definition, it is zero for the fired pixel.
In this measure we have the combination of all possible crosstalk effects we can have in the module: crosstalk between SiPMs (light reflected from the window), between signal lines, between ASIC channels, and between connector pins.
The charge crosstalk is always less than \qty[]{1.4}{\percent}, and peak-to-peak crosstalk is always below \qty[]{5}{\percent}. 
While the first one is important for understanding the quality of the Cherenkov images, the second one is critical for the trigger.
In fact, a trigger signal occurs when the sum of four adjacent pixels (superpixel) exceeds a certain threshold; if there is crosstalk in an adjacent channel, it can negatively affect the trigger. This effect is currently under investigation.
\par
The next batch of modules have been produced and are currently undergoing tests in the University of Leicester, UK, at Max Plank Institute for Nuclear Physics, Germany and at Erlangen Centre for Astroparticle Physics, Germany. Further tests will be critical to verify the initial results reported here.
\par
Regarding the illumination system, a prototype flasher has been extensively tested, demonstrating its ability to produce fast and variable intensity light pulses. The design of the flasher is now being finalized.
A first prototype of the Slow Control Board has been built and successfully tested and will soon be integrated into the door control system.
A complete set of doors have been built and tested. 
The enclosure design is completed and a first version has been built and tested.
%
\section{Conclusions and Future Prospects}
The design of the SST camera is nearly complete and the first copies of subsystems are being tested. Initial results are positive and indicate that all design requirements will be met.
At the end of 2023, the first quarter of the camera (QCAM) is planned to be built. It will consist of 8 camera modules (512 readout channels), the Slow Control Board, all the mechanics and a quarter Backplane.
QCAM will then undergo detailed laboratory testing, after which a complete engineering camera (ECAM) will be built and commissioned on a prototype telescope structure.
%
\section*{Acknowledgments}
This work was conducted in the context of the CTA SST Collaboration.
We are grateful for financial support from organizations and agencies listed in \url{https://www.cta-observatory.org/consortium_acknowledgments/}.
\bibliography{Bibliography}



\clearpage
\section*{Full Authors List: CTA SST Project}
\footnotesize
\raggedright
\mbox{Antonelli Angelo$^{\ref{AFFIL::INAFOARm}}$},
\mbox{Arnesen Tora$^{\ref{AFFIL::NetherlandsUGroningen}}$},
\mbox{Aschersleben Jann$^{\ref{AFFIL::NetherlandsUGroningen}}$},
\mbox{Attinà Primo$^{\ref{AFFIL::ItalyOBrera}}$},
\mbox{Balbo Matteo$^{\ref{AFFIL::SwitzerlandUGenevaISDC}}$},
\mbox{Bang Sunghyun$^{\ref{AFFIL::JapanUNagoya}}$},
\mbox{Barcelo Miquel$^{\ref{AFFIL::GermanyMPIK}}$},
\mbox{Baryshev Andrey$^{\ref{AFFIL::NetherlandsUGroningen}}$},
\mbox{Bellassai Giancarlo$^{\ref{AFFIL::ItalyOCatania}}$},
\mbox{Berge David$^{\ref{AFFIL::GermanyDESY}}$},
\mbox{Bicknell Chris$^{\ref{AFFIL::UnitedKingdomULeicester}}$},
\mbox{Bigongiari Ciro$^{\ref{AFFIL::INAFOARm}}$},
\mbox{Bonnoli Giacomo$^{\ref{AFFIL::ItalyOBrera}}$},
\mbox{Bouley Frederic$^{\ref{AFFIL::FranceObservatoiredeParis}}$},
\mbox{Brown Anthony$^{\ref{AFFIL::UnitedKingdomUDurham}}$},
\mbox{Bulgarelli Andrea$^{\ref{AFFIL::ItalyOASBologna}}$},
\mbox{Cappi Massimo$^{\ref{AFFIL::ItalyOASBologna}}$},
\mbox{Caraveo Patrizia$^{\ref{AFFIL::ItalyIASFMilano}}$},
\mbox{Caschera Salvatore$^{\ref{AFFIL::ItalyOCagliari}}$},
\mbox{Chadwick Paula$^{\ref{AFFIL::UnitedKingdomUDurham}}$},
\mbox{Conte Francesco$^{\ref{AFFIL::GermanyMPIK}}$},
\mbox{Cotter Garret$^{\ref{AFFIL::UnitedKingdomUOxford}}$},
\mbox{Cristofari Pierre$^{\ref{AFFIL::FranceObservatoiredeParis}}$},
\mbox{Dal~Pino Elisabete~Maria~de~Gouveia$^{\ref{AFFIL::BrazilEACHUSaoPaulo}}$},
\mbox{De~Frondat Fatima$^{\ref{AFFIL::FranceObservatoiredeParis}}$},
\mbox{De~Simone Nico$^{\ref{AFFIL::GermanyDESY}}$},
\mbox{Depaoli Davide$^{\ref{AFFIL::GermanyMPIK}}$},
\mbox{Dournaux Jean-Laurent$^{\ref{AFFIL::FranceObservatoiredeParis}}$},
\mbox{Duffy Connor$^{\ref{AFFIL::UnitedKingdomUOxford}}$}
\mbox{Einecke Sabrina$^{\ref{AFFIL::AustraliaUAdelaide}}$},
\mbox{Fermino Carlos~Eduardo$^{\ref{AFFIL::BrazilEACHUSaoPaulo}}$},
\mbox{Funk Stefan$^{\ref{AFFIL::GermanyUErlangenECAP}}$},
\mbox{Gargano Carmelo$^{\ref{AFFIL::ItalyOPalermo}}$},
\mbox{Giavitto Gianluca$^{\ref{AFFIL::GermanyDESY}}$},
\mbox{Giuliani Andrea$^{\ref{AFFIL::ItalyIASFMilano}}$},
\mbox{Greenshaw Tim$^{\ref{AFFIL::UnitedKingdomULiverpool}}$},
\mbox{Hinton Jim$^{\ref{AFFIL::GermanyMPIK}}$},
\mbox{Iovenitti Simone$^{\ref{AFFIL::ItalyOBrera}}$},
\mbox{La~Palombara Nicola$^{\ref{AFFIL::ItalyIASFMilano}}$},
\mbox{Lapington Jon$^{\ref{AFFIL::UnitedKingdomULeicester}}$},
\mbox{Laporte Philippe$^{\ref{AFFIL::FranceObservatoiredeParis}}$},
\mbox{Leach Steve$^{\ref{AFFIL::UnitedKingdomULeicester}}$},
\mbox{Lessio Luigi$^{\ref{AFFIL::ItalyOPadova}}$},
\mbox{Leto Giuseppe$^{\ref{AFFIL::ItalyOCatania}}$},
\mbox{Lloyd Sheridan$^{\ref{AFFIL::UnitedKingdomUDurham}}$},
\mbox{Lombardi Saverio$^{\ref{AFFIL::INAFOARm}}$},
\mbox{Lucarelli Fabrizio$^{\ref{AFFIL::INAFOARm}}$},
\mbox{Macchi Alberto$^{\ref{AFFIL::ItalyOBrera}}$},
\mbox{Martinetti Eugenio$^{\ref{AFFIL::ItalyOCatania}}$},
\mbox{Miccichè Antonio$^{\ref{AFFIL::ItalyOCatania}}$},
\mbox{Millul Rachele$^{\ref{AFFIL::ItalyOBrera}}$},
\mbox{Mineo Teresa$^{\ref{AFFIL::ItalyOPalermo}}$},
\mbox{Mitsunari Takahashi$^{\ref{AFFIL::JapanUNagoya}}$},
\mbox{Nayak Amrit$^{\ref{AFFIL::UnitedKingdomUDurham}}$},
\mbox{Nicotra Gaetano$^{\ref{AFFIL::ItalyRadioastronomiaINAF}}$},
\mbox{Okumura Akira$^{\ref{AFFIL::JapanUNagoya}}$},
\mbox{Pareschi Giovanni$^{\ref{AFFIL::ItalyOBrera}}$},
\mbox{Penno Marek$^{\ref{AFFIL::GermanyDESY}}$},
\mbox{Prokoph Heike$^{\ref{AFFIL::GermanyDESY}}$},
\mbox{Rebert Emma$^{\ref{AFFIL::FranceObservatoiredeParis}}$},
\mbox{Righi Chiara$^{\ref{AFFIL::ItalyOBrera}}$},
\mbox{Rodeghiero Gabriele$^{\ref{AFFIL::ItalyOASBologna}}$},
\mbox{Ross Duncan$^{\ref{AFFIL::UnitedKingdomULeicester}}$},
\mbox{Rowell Gavin$^{\ref{AFFIL::AustraliaUAdelaide}}$},
\mbox{Rulten Cameron$^{\ref{AFFIL::UnitedKingdomUDurham}}$},
\mbox{Russo Federico$^{\ref{AFFIL::ItalyOASBologna}}$},
\mbox{Sanchez~Ricardo Zanmar$^{\ref{AFFIL::ItalyOCatania}}$},
\mbox{Saturni Francesco$^{\ref{AFFIL::INAFOARm}}$},
\mbox{Schaefer Johannes$^{\ref{AFFIL::GermanyUErlangenECAP}}$},
\mbox{Schwab Benni$^{\ref{AFFIL::GermanyUErlangenECAP}}$},
\mbox{Scuderi Salvatore$^{\ref{AFFIL::ItalyIASFMilano}}$},
\mbox{Sironi Giorgia$^{\ref{AFFIL::ItalyOBrera}}$},
\mbox{Sliusar Vitalii$^{\ref{AFFIL::SwitzerlandUGenevaISDC}}$},
\mbox{Sol Helene$^{\ref{AFFIL::FranceObservatoiredeParis}}$},
\mbox{Spencer Samuel$^{\ref{AFFIL::GermanyUErlangenECAP}}$},
\mbox{Stamerra Antonio$^{\ref{AFFIL::INAFOARm}}$},
\mbox{Tagliaferri Gianpiero$^{\ref{AFFIL::ItalyOBrera}}$},
\mbox{Tajima Hiro$^{\ref{AFFIL::JapanUNagoya}}$},
\mbox{Tavecchio Fabrizio$^{\ref{AFFIL::ItalyOBrera}}$},
\mbox{Tosti Gino$^{\ref{AFFIL::ItalyOBrera}}$},
\mbox{Trois Alessio$^{\ref{AFFIL::ItalyOCagliari}}$},
\mbox{Vecchi Manuela$^{\ref{AFFIL::NetherlandsUGroningen}}$},
\mbox{Vercellone Stefano$^{\ref{AFFIL::ItalyOBrera}}$},
\mbox{Vink Jacco$^{\ref{AFFIL::NetherlandsUAmsterdam}}$},
\mbox{Walter Roland$^{\ref{AFFIL::SwitzerlandUGenevaISDC}}$},
\mbox{Watson Jason$^{\ref{AFFIL::GermanyDESY}}$},
\mbox{White Richard$^{\ref{AFFIL::GermanyMPIK}}$},
\mbox{Wohlleben Frederik$^{\ref{AFFIL::GermanyMPIK}}$},
\mbox{Zampieri Luca$^{\ref{AFFIL::ItalyOPadova}}$},
\mbox{Zech Andreas$^{\ref{AFFIL::FranceObservatoiredeParis}}$},
\mbox{Zink Adrian$^{\ref{AFFIL::GermanyUErlangenECAP}}$}
%
\begin{enumerate}[label=$^{\arabic*}$,ref=\arabic*,leftmargin=1em,labelsep=0.25em,labelwidth=1.25em,itemsep=-0.01em]
    \item INAF - Istituto di Astrofisica e Planetologia Spaziali (IAPS), Via del Fosso del Cavaliere 100, 00133 Roma, Italy \label{AFFIL::INAFOARm}
    \item Kapteyn Astronomical Institute, University of Groningen, Landleven 12, 9747 AD, Groningen, The Netherlands\label{AFFIL::NetherlandsUGroningen}
    \item INAF - Osservatorio Astronomico di Brera, Via Brera 28, 20121 Milano, Italy\label{AFFIL::ItalyOBrera}
    \item Department of Astronomy, University of Geneva, Chemin d'Ecogia 16, CH-1290 Versoix, Switzerland \label{AFFIL::SwitzerlandUGenevaISDC}
    \item Department of Physics, Nagoya University, Chikusa-ku, Nagoya, 464-8602, Japan \label{AFFIL::JapanUNagoya}
    \item Max Planck Institute for Nuclear Physics, Saupfercheckweg 1, 69117 Heidelberg, Germany, Saupfercheckweg 1, 69117 Heidelberg, Germany\label{AFFIL::GermanyMPIK}
    \item INAF - Osservatorio Astronomico di Cagliari, Via della Scienza 5, 09047 Selargius, Italy \label{AFFIL::ItalyOCagliari}
    \item Deutsches Elektronen-Synchrotron, Platanenallee 6, 15738 Zeuthen, Germany \label{AFFIL::GermanyDESY}
    \item School of Physics and Astronomy, University of Leicester, Leicester, LE1 7RH, United Kingdom \label{AFFIL::UnitedKingdomULeicester}
    \item LUTH, GEPI and LERMA, Observatoire de Paris, Universit\'e PSL, Universit\'e Paris Cit\'e, CNRS, 5 place Jules Janssen, 92190, Meudon, France \label{AFFIL::FranceObservatoiredeParis}
    \item Centre for Advanced Instrumentation, Department of Physics, Durham University, South Road, Durham, DH1 3LE, United Kingdom \label{AFFIL::UnitedKingdomUDurham}
    \item INAF - Osservatorio di Astrofisica e Scienza dello spazio di Bologna, Via Piero Gobetti 93/3, 40129  Bologna, Italy \label{AFFIL::ItalyOASBologna}
    \item INAF - Istituto di Astrofisica Spaziale e Fisica Cosmica di Milano, Via A. Corti 12, 20133 Milano, Italy \label{AFFIL::ItalyIASFMilano}
    \item INAF - Osservatorio Astrofisico di Catania, Via S. Sofia, 78, 95123 Catania, Italy \label{AFFIL::ItalyOCatania}
    \item University of Oxford, Department of Physics, Clarendon Laboratory, Parks Road, Oxford, OX1 3PU, United Kingdom \label{AFFIL::UnitedKingdomUOxford}
    \item Escola de Artes, Ci\^encias e Humanidades, Universidade de S\~ao Paulo, Rua Arlindo Bettio, CEP 03828-000, 1000 S\~ao Paulo, Brazil \label{AFFIL::BrazilEACHUSaoPaulo}
    \item School of Physics, Chemistry and Earth Sciences, University of Adelaide, Adelaide SA 5005, Australia \label{AFFIL::AustraliaUAdelaide}
    \item Friedrich-Alexander-Universit\"at Erlangen-N\"urnberg, Erlangen Centre for Astroparticle Physics, Nikolaus-Fiebiger-Str. 2, 91058 Erlangen, Germany \label{AFFIL::GermanyUErlangenECAP}
    \item INAF - Osservatorio Astronomico di Palermo {\textquotedblleft}G.S. Vaiana{\textquotedblright}, Piazza del Parlamento 1, 90134 Palermo, Italy \label{AFFIL::ItalyOPalermo}
    \item University of Liverpool, Oliver Lodge Laboratory, Liverpool L69 7ZE, United Kingdom \label{AFFIL::UnitedKingdomULiverpool}
    \item INAF - Osservatorio Astronomico di Padova, Vicolo dell'Osservatorio 5, 35122 Padova, Italy \label{AFFIL::ItalyOPadova}
    \item INAF - Istituto di Radioastronomia, Via Gobetti 101, 40129 Bologna, Italy \label{AFFIL::ItalyRadioastronomiaINAF}
    \item Anton Pannekoek Institute/GRAPPA, University of Amsterdam, Science Park 904 1098 XH Amsterdam, The Netherlands \label{AFFIL::NetherlandsUAmsterdam}

\end{enumerate}

\end{document}